\documentclass[pdflatex,sn-nature]{sn-jnl}
\usepackage{epstopdf}
\usepackage{pdflscape}
\usepackage{adjustbox}
\usepackage{amsmath}   % 提供 aligned 环境, \overset 等高级数学功能
\usepackage{amssymb}   % 提供 \mathbb, \mathcal 等字体符号
\usepackage{mathtools} % amsmath 的扩展，提供更多数学排版修正
\def\be{\begin{equation}}
\def\ee{\end{equation}}
\def\bee{\begin{eqnarray}}
\def\ene{\end{eqnarray}}
\def\bes{\begin{subequations}}
\def\ees{\end{subequations}}

\usepackage[abs]{overpic}  % per sovrapporre testo alle immagini

\usepackage{algorithm}     % 提供浮动体环境 \begin{algorithm}
\usepackage{algorithmic}   % 提供具体的指令环境 \begin{algorithmic}
\usepackage{xcolor}  % 颜色支持
\usepackage{empheq}  % 公式美化支持
\usepackage{graphicx} % figures
\usepackage{hyperref} % hyperlinks
\usepackage{cleveref} % must be loaded after hyperref
\usepackage{booktabs}  % 提供三线表画线命令: \toprule, \midrule, \bottomrule
\usepackage{multirow}  % 提供跨行表格功能: \multirow
\usepackage{amsmath}   % 提供数学公式支持 (align, equation 等)
\usepackage{amssymb}   % 提供更多数学符号 (如 \mathbb, \mathfrak 等)
\usepackage{amsthm}    % 提供定理环境支持 (theorem, proof 等)
\usepackage{mathrsfs}  % 提供花体字母 (如 \mathscr)
\usepackage{bm}        % 提供加粗数学符号 (\bm)
\theoremstyle{plain} % 默认样式（标题加粗，内容斜体）
\usepackage{framed,multirow}
\usepackage[utf8]{inputenc}
\usepackage{booktabs} % 提供更好的表格线段
\usepackage{multirow} % 处理跨行
\usepackage{geometry}
\usepackage{amssymb}
\usepackage{latexsym}
\usepackage{booktabs} % For professional table lines (\toprule, \midrule, \bottomrule)
\usepackage{multirow}

\usepackage[caption=false,justification=raggedleft]{subfig}
\usepackage{etoolbox}
\robustify{\phantom}

\usepackage{url}
\definecolor{newcolor}{rgb}{.8,.349,.1}



\renewcommand{\proofname}{Proof} % 英文

\begin{document}

% \verso{Given-name Surname \textit{etal}}

\title{Cluster Attention Neural Operators for Solving Parametric Partial Differential Equations}
% \tnotetext[tnote1]{This is an example for title footnote coding.}

%% --- 作者部分 ---

%% 第一作者: Ming Zhong (机构 1, 2)
\author[1,2,3]{\fnm{Ming} \sur{Zhong}}

%% 第二作者: Lu Lu (机构 3, 4)
\author[3]{\fnm{Antonio} \sur{Colanera}}

%% 第三作者: Weifang Weng (机构 5)
\author[3]{\fnm{Gianluigi} \sur{Rozza}}

%% 第四作者(通讯作者): Zhenya Yan (机构 6, 2)
\author[4,2,1]{\fnm{Zhenya} \sur{Yan}}\email{zyyan@mmrc.iss.ac.cn (Z. Yan)}

%% --- 通讯作者角标说明 ---

%% --- 机构列表 (对应上面的数字) ---

\affil[1]{\it School of Advanced Interdisciplinary Sciences, University of Chinese Academy of Sciences, Beijing 100049, China}

\affil[2]{\it State Key Laboratory of Mathematical Sciences, Academy of Mathematics and Systems Science, Chinese Academy of Sciences, Beijing 100190, China}

\affil[3]{\it International School for Advanced Studies (SISSA), Trieste 34136, Italy}

\affil[4]{\it School of Mathematics and Information Science, Zhongyuan University of Technology, Zhengzhou 450007, China}

%\affil[5]{School of Mathematical Sciences, University of Chinese Academy of Sciences, Beijing 100049, China}
% \received{1 May 2013}
% \finalform{10 May 2013}
% \accepted{13 May 2013}
% \availableonline{15 May 2013}
% \communicated{S. Sarkar}

\vspace{-0.2in}
%%%
\abstract{
Traditional simulations of parametric partial differential equations (PDEs) rely on repetitive computations for each parameter, which makes high-fidelity design  impractical. Neural operators address  this issue by learning solution operators, accelerating parameter-space mapping by orders of magnitude. Recent Transformer-based neural operators attempt to capture global dependencies, but often at the cost of quadratic attention complexity. Transolver  resolves this problem by projecting physical states into a reduced slice space for attention computation. Although fast, this projection sacrifices fine spatial information. Moreover, by operating  in this reduced space with shared weights across attention heads, it may constrain the model's  flexibility, thereby limiting its capacity to  capture complex phenomena. To address these issues, we propose the Cluster Attention Neural Operator (CANO), which reformulates attention via a novel cross-attention mechanism that dynamically clusters queries while preserving full-resolution keys and values. This avoids slice compression loss and removes weight-sharing limits. At the same time, the model remains fast without losing global interactions. Empirically, CANO achieves state-of-the-art performance across canonical PDE benchmarks, covering fluid and solid dynamics (e.g., Navier-Stokes, Airfoil, Plasticity), irregular unstructured geometries (e.g., Pipe Turbulence, Composites), and long-term temporal rollouts. Across solid deformation and turbulent flow benchmarks, CANO achieves lower errors than baselines and exhibits strong geometric adaptability and temporal consistency.
}


\maketitle

\section{Introduction}
\label{IN}

Partial differential equations (PDEs) play a pivotal role in   modeling
physical phenomena and have became ubiquitous  across science and engineering~\cite{friedman1975stochastic,braun1983differential,smith2010introduction,logan2014applied,simmons2016differential}. While these PDEs provide a  theoretical framework for understanding natural phenomena, yet solving these equations numerically is often too slow for practical use~\cite{jaun1999numerical,num00,wendt2008computational,tadmor2012review,ames2014numerical}. Traditional numerical solvers, such as the finite element, finite difference, and spectral methods, typically rely on dense spatial discretizations \cite{jaun1999numerical,num00,wendt2008computational,tadmor2012review,ames2014numerical}. However, the computational cost of standard methods grows steeply with grid resolution. As a result, high-fidelity simulations are often computationally infeasible. This leaves a significant gap between theoretical models and real-world engineering \cite{han2017,Jeon2025}.

 Recently, scientific machine learning (SciML) has emerged as a promising alternative, replacing classical iterative solvers with trained neural networks~\cite{dl1,dl2,dl3,dl4}. For example, the Deep Ritz method \cite{yu2018deep} solves PDEs via variational formulations, while Physics-Informed Neural Networks (PINNs) \cite{raissi2019physics, lu2021deepxde, pinn} penalize PDE residuals in the loss function. Other variants, such as PeRCNN \cite{rao2023encoding}, enforce physical laws directly through network architectures.
However,  these methods only solve one PDE instance at a time. Once the physical parameters or boundary conditions changed, the network must be re-trained from scratch, making it impractical for real-time applications\cite{krishnapriyan2021characterizing, wang2022and}.

To address this issues, neural operators (NOs) \cite{kovachki2021universal,li2020fourier,lu2021learning,kovachki2023neural,lanthaler2025nonlocality} learn mappings  between   parameter or coefficient fields to solution fields. Instead of solving a single PDE instance, they approximate the underlying operator. Once trained, they provide zero-shot solution predictions for unseen configurations at a fraction of the classical computational cost.

 Existing neural operator architectures can be broadly categorized into MLP-based, spectral-based, and transformer-based methods.
 % \begin{itemize}
 %     \item
 First, MLP-based neural operators, represented by Deep Operator Networks (DeepONet) \cite{lu2021learning}, are fundamentally grounded in the universal approximation theorem for operators \cite{chen1995universal}. This framework has been extended through various architectures, including POD-DeepONet \cite{lu2022pod}, PINN-DeepONet \cite{wang2021learning}, MIONet \cite{jin2022mionet}, and Hybrid-preconditioned DeepONet \cite{kopanivcakova2025deeponet}. In practice, however, these simple architectures struggle to resolve complex nonlinear behaviors.

 % \item
 Second, spectral-based neural operators, such as the Fourier Neural Operator (FNO) \cite{li2020fourier,kovachki2021universal,li2023fourier,li2023geometry}, exploit global convolutions in the wavenumbers domain to parameterize integral kernels, achieving high accuracy on structured grids. However, accommodating general unstructured domains significantly increases computational cost or requires lossy interpolations \cite{li2023fourier,li2023geometry}.

% \item
Third, Transformer-based neural operators \cite{cao2021choose,li2022transformer,hao2023gnot,li2023scalable,xiao2023improved,shih2025transformers} offer an alternative way to handle unstructured domains and irregular meshes. A standard self-attention block computes the output $ \mathbf{Y}$ as \cite{vaswani2017attention,devlin2019bert,brown2020language}
 \begin{equation}\label{soft}
 \mathbf{Y} = \mathrm{Softmax}\Big(\frac{\mathbf{Q}\mathbf{K}^\top}{\sqrt{d}}\Big)\mathbf{V},
 \end{equation}
 where $\mathbf{Q}, \mathbf{K}, \mathbf{V} \in \mathbb{R}^{N \times d}$ are the query, key, and value matrices, respectively, obtained from the projection of original features (such as function values or coordinates), where $N$ is the number of tokens and $d$ is the feature dimension.  While this formulation captures long-range dependencies, it scales quadratically, $\mathcal{O}(N^2 d)$, making it impractical for fine physical discretizations. Linear attention variants reduce complexity to $\mathcal{O}(N d^2)$ through matrix factorization \cite{katharopoulos2020transformers,wang2020linformer,shen2021efficient,han2024agent,qs24}:
\begin{equation}\label{linear}
 \mathbf{Y} = \phi(\mathbf{Q})\left(\phi(\mathbf{K})^\top \mathbf{V}\right),
 \end{equation}
 where $\phi(\cdot)$ denotes a feature mapping.
 Despite its efficiency, linear attention often suffers from overly smooth attention maps. While hierarchical or convolutional designs \cite{ovadia2024vito,wang2024cvit,liu2024mitigating,dosovitskiy2020image,liu2021swin} mitigate this issue, they are strictly tailored for structured grids and do not readily apply to unstructured meshes.

 % \end{itemize}

 Another line of work tackles complex geometries by introducing latent tokens \cite{wang2024latent,wu2024transolver,alkin2024universal}. A notable example is Transolver \cite{wu2024transolver}, which learns a weight matrix to aggregate input features into a compact "slice space", performs attention within this reduced representation, and projects back. However, compressing physical points into slices loses local geometric details. In addition, Transolver computes attention only in this slice space and shares weights across attention heads, which limits its ability to capture complex dynamics.

 To overcome these challenges, we introduce the {\it Cluster Attention Neural Operator} (CANO)\footnote{\url{https://github.com/zhongming12342/CANO}}. Rather than projecting points into a reduced slice space, CANO computes cross-attention by clustering queries ($Q$) dynamically, while retaining keys ($K$) and values ($V$) in their original coordinate space. By clustering queries dynamically, CANO avoids  compression loss in the slice space and removes the need for weight sharing across attention heads. As a result, the model maintains global receptive fields while scaling linearly with the number of points $N$, yielding strong generalization on complex  physical systems.

The main contributions of this work are twofold. First, we introduce Cluster Attention, which clusters queries while keeping keys and values at full resolution. This formulation prevents detail loss and scales linearly with the number of points $N$. Second, we validate CANO on 12 diverse PDE benchmarks, including complex geometries and long rollout horizons, where it achieves up to 41.67\% lower error on solid mechanics and 33.73\% on turbulent flows compared to previous models.

%  The main contributions of this work are summarized as follows:

% \begin{itemize}
%     \item \textbf{Cluster Attention Mechanism:}
%     We propose an asymmetric cross-attention mechanism that dynamically clusters the query space while maintaining full-resolution keys and values. This design fundamentally removes the information bottleneck inherent in symmetric latent compression, ensuring the high-fidelity reconstruction of multi-scale physical dynamics with linear $\mathcal{O}(N)$ computational complexity.

%     \item \textbf{Extensive Empirical Validation:}
%     We conduct a comprehensive evaluation across 12 challenging PDE benchmarks, encompassing standard physical dynamics, irregular unstructured geometries, and demanding long-time autoregressive rollouts. Experimental results demonstrate that CANO consistently establishes a new state-of-the-art, achieving significant error reductions—such as a 41.67\% improvement on complex solid deformation and 33.73\% on turbulent fluid flows—thereby validating its robust geometric adaptability and superior temporal consistency.
% \end{itemize}

% \vspace{0.1in}

The remainder of this paper is organized as follows.
Section~\ref{NOS} briefly introduces the neural operator framework and related works.
Section~\ref{CANO} presents the proposed  CANO architecture.
Section~\ref{Num} reports extensive numerical experimental results on diverse benchmarks and model analysis, followed by concluding remarks in Section~\ref{sec:conclusions}.

\section{\bf Related works}

\label{NOS}

% Neural operators aim to learn mappings from parameter spaces to solution spaces of partial differential equations (PDEs) in a data-driven manner \cite{li2020fourier,lu2021learning,kovachki2023neural,lanthaler2025nonlocality}, bypassing the need for traditional numerical solvers.

In scientific machine learning, neural operators  aim to learn the  mappings between  parameter spaces and corresponding PDE solution spaces~\cite{li2020fourier,lu2021learning,kovachki2023neural,lanthaler2025nonlocality}. Rather than replacing classical solvers, these models act as fast surrogates that accelerate  inference.

From a mathematical perspective, we consider a generalized PDE formulated over a bounded spatial domain $\Omega \subset \mathbb{R}^{d_x}$. Consider a physical system parameterized by an input field $a(x)$, which can represent  coefficients, initial/boundary conditions, or source terms. The governing equation can be written as
\begin{equation}
\left\{\begin{aligned}
(L_a u)(x) = f(x), & \quad x \in \Omega\subset \mathbb{R}^{d_x},\\ u(x) = u_0(x), & \quad x \in \partial \Omega,
\end{aligned}\right.
\end{equation}
where $f(x)$ denotes the source term. Let $\mathcal{A}$ and $\mathcal{U}$ be Banach spaces containing the input parameter field $a$ and the target solution $u: \Omega \to \mathbb{R}$, respectively. Given a parametric differential operator $\mathcal{L}_a: \mathcal{U} \to \mathcal{U}^*$, the goal of operator learning is to approximate the underlying solution operator $\mathcal{G}^\dagger: \mathcal{A} \to \mathcal{U}$ such that $\mathcal{G}^\dagger(a) = u$.

Given a training dataset of input--solution pairs $\{(a^{(i)}, u^{(i)})\}_{i=1}^M$, we approximate $\mathcal{G}^\dagger$ using a parameterized neural operator $\mathcal{G}_\theta: \mathcal{A} \to \mathcal{U}$ with weights $\theta \in \mathbb{R}^p$. The optimal parameters $\theta^*$ are obtained by minimizing the empirical mean squared error in the Bochner norm~\cite{kovachki2023neural}.
% \begin{equation}\left|G^\dagger - G_\theta\right|{L^2\mu(A; U)}^2:= \mathbb{E}{a \sim \mu} \left[\left|G^\dagger(a) - G\theta(a)\right|U^2 \right].
% \end{equation}
In practice, the risk is approximated by the empirical risk over the training samples:
\begin{equation}
\min_{\theta \in \mathbb{R}^p} \frac{1}{M} \sum_{i=1}^M \left\| u^{(i)} - \mathcal{G}_\theta(a^{(i)}) \right\|_{\mathcal{U}}^2.
\end{equation}
Within this framework, neural operator architectures such as the Graph Kernel Network (GKN) \cite{anandkumar2020neural,li2020multipole} are closely motivated by Green's functions. For a linear differential operator $\mathcal{L}_a$ with Green's function $G(x, y)$ satisfying $\mathcal{L}_a G(x, \cdot) = \delta_x$, the solution can be written as an integral convolution:
\begin{equation}
u(x) = \int_\Omega G(x, y) f(y) \, dy.
\end{equation}
Analogously, iterative neural operators update a continuous hidden representation $v_t(x)$ through non-local kernel integration:
\begin{equation}\label{IT}
v_{t+1}(x) = \sigma \Bigg( W v_t(x) + \int_\Omega \kappa_\theta\big(x, y, a(x), a(y)\big) v_t(y) \, \nu_x(dy) \Bigg).
\end{equation}
Here, an initial lifting layer $P$ maps the local input $(x, a(x))$ to the hidden state $v_0(x) = P(x, a(x))$. Each update layer combines a local linear transformation parameterized by weight matrix $W$ with global kernel aggregation via the parameterized kernel $\kappa_\theta$, followed by an activation function $\sigma$ and integration measure $\nu_x$. After $T$ iterative updates, a projection layer $Q$ decodes the final hidden state to obtain the solution field $u(x) = Q(v_T(x))$. This formulation allows the operator to capture both local features and long-range dependencies across the domain.

 To realize the operator learning framework described above, three of the most representative architectures are: the Fourier Neural Operator (FNO) \cite{li2020fourier,kovachki2021universal}, the Deep Operator Network (DeepONet) \cite{lu2021learning}, and Transformer-based neural operators \cite{cao2021choose,li2022transformer,wu2024transolver}.  The FNO \cite{li2020fourier,kovachki2021universal} accelerates non-local integration by assuming a translation-invariant kernel, $\kappa(x, y) = \kappa(x - y)$. Under this assumption, the integral operator reduces to a spatial convolution, which can be evaluated efficiently in the frequency domain via the convolution theorem:
\begin{equation}
v_{t+1}(x) = \sigma \Big( \mathcal{F}^{-1} \big( \widehat{\kappa}_\theta \cdot \mathcal{F}(v_t) \big)(x) + W v_t(x) \Big),
\end{equation}
where $\mathcal{F}$ and $\mathcal{F}^{-1}$ denote the forward and inverse Fourier transforms, and $\widehat{\kappa}_\theta$ is a learnable Fourier multiplier. Truncating the spectrum to the lowest $k_{\text{max}}$ modes allows FNO to run in $\mathcal{O}(N \log N)$ time using the Fast Fourier Transform (FFT). However, because standard FFTs require uniform Cartesian grids and periodic boundaries, standard FNO is difficult to apply directly to complex, unstructured meshes.

An alternative  approach is  DeepONet \cite{lu2021learning,jin2022mionet}, which approximates the mapping $G^\dagger$ based on the universal approximation theorem for operators \cite{chen1995universal}. DeepONet decomposes the solution into a dot product of a branch network and a trunk network:\begin{equation}
G_\theta(a)(x) = \sum_{k=1}^{m} b_k(a) t_k(x),
\end{equation}where $b_k(a)$ are outputs of the branch network encoding the input parameter function $a$, and $t_k(x)$ are outputs of the trunk network encoding the spatial coordinates $x$, $m$ is a pre-specified parameter. While DeepONet offers significant mathematical generality, its performance is often constrained by the learning dynamics of its underlying multi-layer perceptrons (MLPs). Specifically, these architectures exhibit spectral bias \cite{rahaman2019spectral}, meaning they efficiently learn low-frequency macroscopic trends but struggle to resolve the sharp, high-frequency oscillations that are critical for accurate PDE modeling.

To handle unstructured domains and capture complex long-range dependencies, attention mechanisms \cite{vaswani2017attention} have been increasingly adopted for PDE operators. Given a set of latent representations $\mathbf{X} \in \mathbb{R}^{N \times d}$, standard transformer blocks project these into query, key, and value matrices:
\begin{equation}\label{eq:project}
  \mathbf{Q} = \mathbf{X} \mathbf{W}_Q, \quad
\mathbf{K} = \mathbf{X} \mathbf{W}_K,\quad \mathbf{V} = \mathbf{X} \mathbf{W}_V,
\end{equation}
where $\mathbf{W}_Q, \mathbf{W}_K, \mathbf{W}_V \in \mathbb{R}^{d \times d}$. The standard Softmax attention computes the output as displayed in Eq.~(\ref{soft}).
While this formulation effectively models global pairwise interactions, it requires computing and storing the full $N \times N$ similarity matrix,  with a quadratic computational cost of $\mathcal{O}(N^2 d)$. Linear attention methods \cite{katharopoulos2020transformers,wang2020linformer,shen2021efficient,han2024agent} reduce this complexity via a non-negative feature mapping function $\phi(\cdot)$ to decompose the kernel. By leveraging the associativity of matrix multiplication, the operation is rewritten as displayed in Eq.~(\ref{linear}).
 This rearrangement evaluates the $d \times d$ context matrix first,  avoiding the $N \times N$ attention matrix and reducing the complexity to $\mathcal{O}(N d^2)$. However, this  factorization leads to low-rank, smooth attention maps, which limits the model's ability to capture high-frequency details.

To avoid both the quadratic cost of softmax attention and the over-smoothing of linear attention, recent models compress physical tokens into a smaller set of latent representations. For example, Transolver \cite{wu2024transolver} uses a physics-aware slicing mechanism to project the mesh into a reduced number of slice tokens. Specifically, given the input physical node features $\mathbf{X} \in \mathbb{R}^{N \times d}$, Transolver learns a normalized assignment matrix $\mathbf{A} \in \mathbb{R}^{N \times M}$ to partition the domain into $M$ slices, where $M \ll N$. The model aggregates the high-dimensional node features into a compact "slice space" via a projection:
\begin{equation}
\mathbf{S} = \mathbf{A}^\top \mathbf{X}, \qquad \mathbf{S} \in \mathbb{R}^{M \times d}.
\end{equation}
Standard Softmax attention is then executed exclusively on these $M$ slice tokens to capture global interactions efficiently:
\begin{equation}
\mathbf{S}' = \mathrm{Softmax}\Big(\frac{(\mathbf{S} \mathbf{W}_Q)(\mathbf{S} \mathbf{W}_K)^\top}{\sqrt{d}}\Big) (\mathbf{S}\mathbf{W}_V),
\end{equation}
where $\mathbf{W}_Q, \mathbf{W}_K, \mathbf{W}_V \in \mathbb{R}^{d \times d}$ are the learnable projection matrices. Note that in the implementation of Transolver\footnote{\url{https://github.com/thuml/Transolver/blob/main/PDE-Solving-StandardBenchmark/model/Physics_Attention.py}}, these matrices are defined as $\in \mathbb{R}^{d_\text{head} \times d_\text{head}}$, forcing weight sharing across different attention heads. Here $d_\text{head}$ denotes the embedding dimension per head, which is typically calculated as $d / H$ with $H$ being the number of attention heads. Finally, a reverse projection (deslicing) is applied to broadcast the updated slice features back to the original $N$ physical nodes:\begin{equation}\mathbf{Y} = \mathbf{A} \mathbf{S}', \quad \mathbf{Y} \in \mathbb{R}^{N \times d}.\end{equation}
Although this slice-and-broadcast design achieves $\mathcal{O}(N M d)$ linear complexity, it has notable limitations.

Specifically, constructing all queries, keys, and values from compressed slices discards local high-frequency details. In addition, Transolver shares attention weights across heads, which prevents individual heads from learning distinct physical patterns. To resolve these issues, our cluster attention mechanism preserves full-resolution keys and values, clustering only the queries dynamically. This allows unconstrained attention heads to capture  physical interactions while maintaining linear $\mathcal{O}(N)$ complexity.

\section{Cluster Attention Neural Operator (CANO)}
\label{CANO}
As discussed in the previous sections, existing latent compression methods like Transolver cut fine-grained physical structures by projecting the input $\mathbf{X}$ into a highly compressed slice space. In such symmetric compression process, all three matrices including $Q$, $K$, and $V$, suffer from slice compression,  which limits the model's performance on complex problems.

To resolve this  issue, we propose the Cluster Attention Neural Operator (CANO). The key innovation of CANO is the asymmetric cross-attention mechanism. Instead of  aggregating the raw input $\mathbf{X}$, we selectively cluster only the query matrix $\mathbf{Q}$, while preserving the keys $\mathbf{K}$ and values $\mathbf{V}$ at their full, uncompressed physical resolution. This achieves linear complexity $\mathcal{O}(N)$ while retaining global context across the domain.

% using linear transformations:
% \begin{equation}
% \mathbf{Q} = \mathbf{X}\mathbf{W}_Q, \quad \mathbf{K} = \mathbf{X}\mathbf{W}_K, \quad \mathbf{V} = \mathbf{X}\mathbf{W}_V,
% \end{equation}
% \ac{same as equation 9?}
% where $\mathbf{W}_Q, \mathbf{W}_K, \mathbf{W}_V \in \mathbb{R}^{d \times d}$.
Specifically, given the input representation $\mathbf{X} \in \mathbb{R}^{N \times d}$, we first project it into the standard query, key, and value spaces as done in Eq.~(\ref{eq:project}).
Unlike Transolver that learns a slice matrix from $\mathbf{X}$, CANO derives a dynamic cluster assignment matrix $\mathbf{C} \in \mathbb{R}^{N \times K_c}$  from the query matrix $\mathbf{Q}$, where $K_c$ is the predefined number of clusters ($K_c \ll N$). This matrix maps the $N$ physical nodes to $K_c$ cluster centroids, and it's  computed as
\begin{equation}
  \mathbf{C} = \mathrm{Softmax}\big((\mathbf{X}\mathbf{W}_Q)\mathbf{W}_c\big),
\end{equation}
where the Softmax operator is applied across the cluster dimension, and $\mathbf{W}_c \in \mathbb{R}^{d \times K_c}$ is the learnable cluster projection matrix.
Using $\mathbf{C}$, we aggregate the query matrix into a compact set of clustered queries $\mathbf{Q}_c$:
\begin{equation}
\mathbf{Q}_c = \mathbf{C}^\top \mathbf{Q}, \quad \mathbf{Q}_c \in \mathbb{R}^{K_c \times d}.
\end{equation}

With the clustered queries established, CANO executes a  efficient cross-attention mechanism. The compact $\mathbf{Q}_c$ attends to the full, uncompressed key matrix $\mathbf{K}$ to extract relevant features from the  value matrix $\mathbf{V}$:
\begin{equation}
\mathbf{Z}_c = \mathrm{Softmax}\Big(\frac{\mathbf{Q}_c \mathbf{K}^\top}{\sqrt{d}}\Big) \mathbf{V}, \quad \mathbf{Z}_c \in \mathbb{R}^{K_c \times d}.
\end{equation}
This asymmetric operation is fundamentally different from Transolver. By avoiding the compression of $\mathbf{K}$ and $\mathbf{V}$, the attention matrix inherently retains physical details.

\begin{figure}
  \centering
\includegraphics[width=1.0\textwidth]{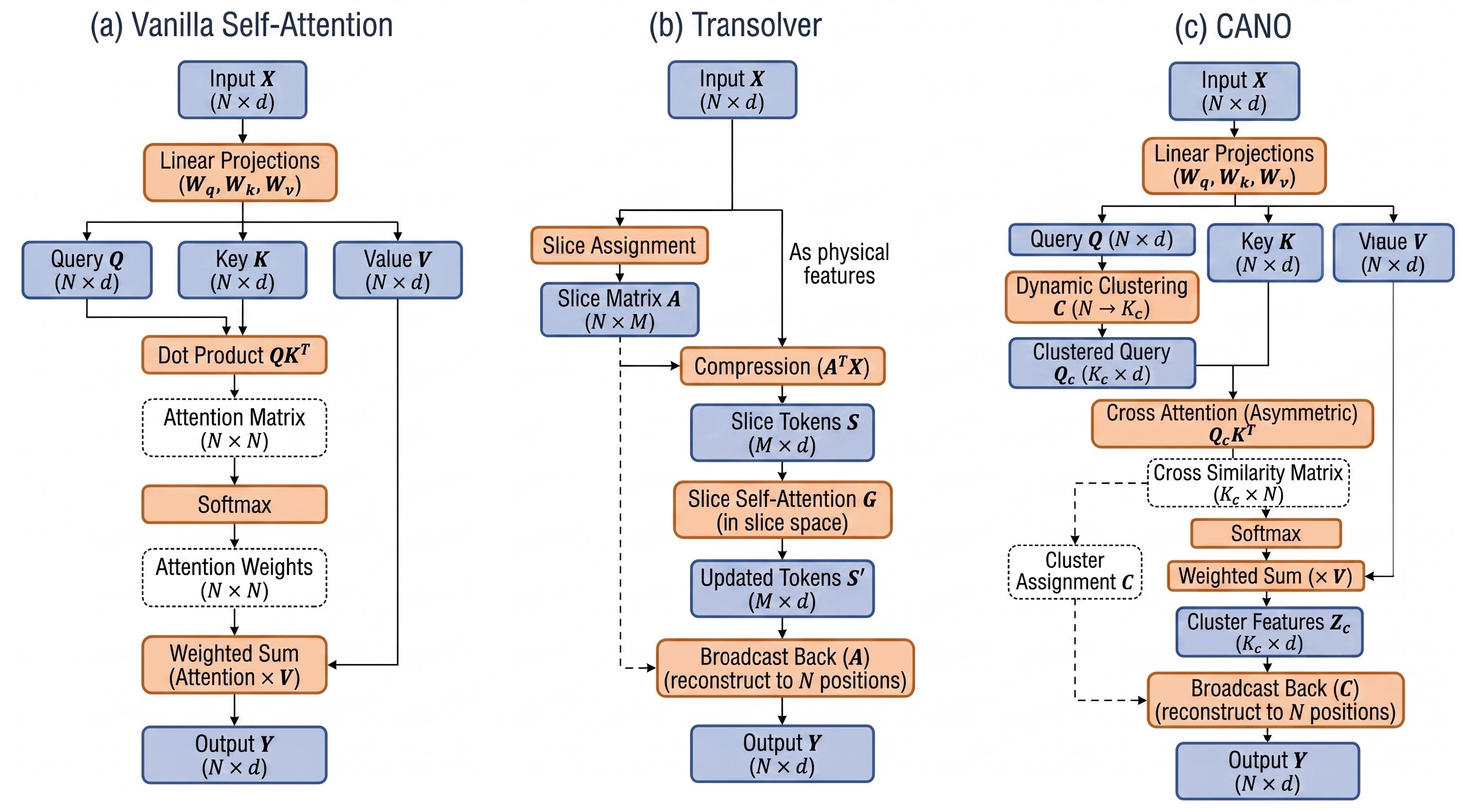}
\vspace{-0.2in}\caption{Comparison of the architecture  of \textbf{(a)} Vanilla Self-Attention, \textbf{(b)} Transolver, and \textbf{(c)} CANO mechanisms.}
\label{Sketch}
\end{figure}

Finally, to reconstruct the updated representations for each individual physical node, we utilize the exact same assignment matrix $\mathbf{C}$ to dispatch (broadcast) the aggregated cluster features $\mathbf{Z}_c$ back to the original spatial resolution:
\begin{equation}
\mathbf{Y} = \mathbf{C} \mathbf{Z}_c, \quad \mathbf{Y} \in \mathbb{R}^{N \times d}.
\end{equation}
A comparison of different attention mechanisms (including vanilla  self-attention, Transolver, and CANO) can be found in Figure ~\ref{Sketch}.

% \ac{I would put here the definition of $\mathbf{C} = \mathrm{Softmax}\big((\mathbf{X}\mathbf{W}_Q)\mathbf{W}_c\big), ...$, in a begin-equation, I think this is an important euqation for the paper }

From a computational perspective, the calculation of the cross-attention matrix $\mathbf{Q}_c \mathbf{K}^\top$ and the subsequent value aggregation require $\mathcal{O}(N K_c d)$ operations. Since $K_c \ll N$, CANO elegantly achieves linear complexity with respect to the sequence length $N$. Furthermore, because its mechanism, CANO inherently supports independent multi-head attention computations without the need to share weights across heads, avoiding the detail loss seen in Transolver. The theoretical connection between CANO and generalized linear attention~\cite{2026linearno} can be seen from Appendix~\ref{link}. The detailed algorithm flow is given in the algorithm \ref{alg:cano_algorithmic_main},  where $\text{FFN}(\cdot)$ and $\text{LayerNorm}(\cdot)$ denote the feed-forward network and layer normalization, respectively.

\begin{algorithm}[!t]
\caption{Cluster Attention Neural Operator (CANO) for solving PDEs}
\label{alg:cano_algorithmic_main}
\begin{algorithmic}[1]
\REQUIRE Spatial mesh coordinates $\mathbf{x} \in \mathbb{R}^{N \times d_{\mathbf{x}}}$, input function values $a(\mathbf{x})  \in \mathbb{R}^{N \times d_{\mathbf{a}}}$, ground truth $u_{\mathrm{gt}}(\mathbf{x}) \in \mathbb{R}^{N \times d_{\mathbf{u}}}$
\ENSURE Predicted solution field $\hat{u}(\mathbf{x}) \in \mathbb{R}^{N \times d_{\mathrm{u}}}$

\STATE Initialize network parameters $\mathcal{W}$, base learning rate $\alpha_0$, total epochs $T_{\mathrm{epochs}}$;
\FOR{$\text{epoch} = 1$ to $T_{\mathrm{epochs}}$}
    \STATE \textbf{Input encoding:}
    \STATE $\mathbf{h}^{(0)} \leftarrow \mathrm{Encoder}([\mathbf{x},\, a(\mathbf{x})]) \in \mathbb{R}^{N \times d}$;
    \STATE \textbf{Cluster Attention Blocks:}
    \FOR{$l = 0$ to $L-1$}
    \STATE $\widetilde{\mathbf{h}}^{(l)} \leftarrow \mathbf{h}^{(l)} + \mathrm{Cluster}\text{-}\mathrm{Attention}(\mathrm{LayerNorm}(\mathbf{H}^{(l)}))$;
    \STATE $\mathbf{h}^{(l+1)} \leftarrow \widetilde{\mathbf{h}}^{(l)} + \mathrm{FFN}(\mathrm{LayerNorm}(\widetilde{\mathbf{h}}^{(l)}))$;
\ENDFOR
    \STATE \textbf{Output decoding:}
    \STATE $\hat{u}(\mathbf{x}) \leftarrow \mathrm{Decoder}(\mathbf{h}^{(L)}) \in \mathbb{R}^{N \times d_{\mathrm{u}}}$;
    \STATE \textbf{Loss Computation:}
    \STATE $\mathcal{L}_{\mathrm{train}} \leftarrow
    \dfrac{\|\hat{u}(\mathbf{x}) - u_{\mathrm{gt}}(\mathbf{x})\|_{L^2}}
    {\|u_{\mathrm{gt}}(\mathbf{x})\|_{L^2}}$;
    \STATE Backpropagate to compute parameter gradients $\nabla_{\mathcal{W}} \mathcal{L}_{\mathrm{train}}$;
    \STATE Adjust current learning rate: $\alpha_{\mathrm{epoch}} \leftarrow \mathrm{Schedule}(\text{epoch}, \alpha_0, T_{\mathrm{epochs}})$;
    \STATE Update  parameters via AdamW:
    \STATE $\mathcal{W} \leftarrow \mathrm{AdamW}(\mathcal{W}, \nabla_{\mathcal{W}} \mathcal{L}_{\mathrm{train}}, \alpha_{\mathrm{epoch}})$;
\ENDFOR
\RETURN $\hat{u}(\mathbf{x})$
\end{algorithmic}
\end{algorithm}

\section{Numerical Experiment Results}
\label{Num}
In this section, we provide a comprehensive evaluation of the proposed CANO from four aspects. First, we compare CANO with state-of-the-art baselines on standard PDE benchmarks~\cite{li2020fourier,li2023fourier}. We then assess its ability to handle complex geometries through experiments on irregular-domain benchmarks~\cite{chen2023}. Next, we investigate its temporal stability in challenging long-horizon autoregressive rollout settings~\cite{li2020fourier,liu2025}. Finally, we analyze the impact of key hyperparameters on the performance of CANO.

% ... existing code ...
To validate the performances of our CANO approach, we compare it against different neural PDE solvers. This includes graph neural networks (e.g., GraphSAGE \cite{hamilton2017}), different  neural operators, ranging from well known architectures like FNO \cite{li2020fourier} and DeepONet \cite{lu2021learning} to their  variants (POD-DeepONet \cite{lu2022pod}, Geo-FNO \cite{li2023fourier}, F-FNO \cite{tran2021factorized}, U-FNO \cite{wen2022u}, and U-NO \cite{rahman2022u}), as well as recent innovations like LSM \cite{wu2023solving}, NORM \cite{chen2023}, and HPM \cite{yue2024}. Furthermore, we evaluate against state-of-the-art Transformer-based operators, including Galerkin \cite{cao2021choose}, OFormer \cite{li2022transformer}, FactFormer \cite{li2023scalable}, GNOT \cite{hao2023gnot}, ONO \cite{xiao2023improved}, HT-Net \cite{liu2024mitigating}, and Transolver \cite{wu2024transolver}. As the strongest baseline, Transolver serves as our primary point of comparison throughout the experiments.

For most baseline methods, we report the performance values provided in the corresponding original publications or in the comprehensive benchmark of the Transolver study. To enable a controlled comparison of different attention mechanisms, we additionally re-implemented the Galerkin Transformer and Transolver within the same training framework adopted for CANO.
 Table \ref{tab:hyperparams} outlines the experimental settings and hyperparameter choices, including cluster numbers ($K_c$) and transformer configurations, used for each benchmark.
 These settings have been selected empirically to provide a suitable trade-off between predictive accuracy and computational cost. For Transolver, the MLP expansion ratio ($r_{\text{mlp}}$ in Table~\ref{tab:hyperparams}) was set to 2 for most benchmarks.

\subsection{Standard PDE Benchmarks}

To ensure a fair comparison against all baseline methods, we have conducted experiments on six standard PDE benchmarks: Airfoil, Pipe, Plasticity, Navier-Stokes, Darcy flow, and Elasticity, as detailed in Table \ref{tab:dataset_overview}.
\begin{table}[!t]
  \centering
  \small
  \setlength{\tabcolsep}{4pt}
  \caption{Unified training and model hyperparameters across benchmarks. All experiments utilize an initial learning rate of $10^{-3}$. The model configuration is denoted as ($L$/$H$/$d$/$r_{\text{mlp}}$), representing layer, number of attention head, embedding dimension and MLP ratio in the transformer blocks.}
  \vspace{2mm}
  \begin{tabular}{l l c c c c}
    \toprule
    \textbf{Benchmark} & \textbf{Loss Function} & \textbf{Epochs} & \textbf{Batch} & \textbf{Cluster Number} & \textbf{$L$/$H$/$d$/$r_\text{mlp}$} \\
    \midrule

    % --- Group 1 ---
    \multicolumn{6}{c}{\textit{Standard benchmarks}} \\
    \midrule
    Airfoil            & Rel.\ $L^2$                 & 500 & 4 & 32 & 8 / 8 / 128 / 2\\
    Pipe               & Rel.\ $L^2$                 & 500 & 4 & 16 & 8 / 4 / 128 / 2\\
     Plasticity         & Rel.\ $L^2$                 & 500 & 8 & 64 & 8 / 8 / 128 / 2\\

    Navier-Stokes      & Rel.\ $L^2$                 & 500 & 2 & 32 & 8 / 8 / 256 / 1\\
    Darcy              & Rel.\ $L^2$ + 0.1$L_\nabla$ & 500 & 4 & 128 & 8 / 8 / 128 / 2\\
     Elasticity         & Rel.\ $L^2$                 & 500 & 1 & 128   & 8 / 8 / 128 / 2 \\
    \midrule

    % --- Group 2 ---
    \multicolumn{6}{c}{\textit{Irregular domain benchmarks }} \\
    \midrule
    Irregular Darcy      & Rel.\ $L^2$             & 2000 & 16 & 128 & 4 / 4 / 64 /4 \\
    Pipe Turbulence           & Rel.\ $L^2$                 & 2000 & 16 & 32 & 4 / 4 / 64 /4 \\
    Composite           & Rel.\ $L^2$                 & 2000 & 16 & 32 & 4 / 4 / 64 / 4 \\
  \midrule
        \multicolumn{6}{c}{\textit{ Long-time rollout}} \\
    \midrule
     Navier-Stokes      & Rel.\ $L^2$             & 500 & 2 & 32 & 8 / 8 / 256 / 1 \\
    ICP Plasma           & Rel.\ $L^2$                 & 1000 & 1 & 32 & 8 / 8 / 256 / 1\\
    Heat Flow           & Rel.\ $L^2$                 & 1000 & 1 & 16 & 8 / 8 / 256 / 1\\
    \bottomrule
  \end{tabular}
  \label{tab:hyperparams}
\end{table}

\begin{table}[htbp]
  \centering
  \small % 保持字号缩小，让宽表格显得精致
  \renewcommand{\arraystretch}{1.2} % 增加行高，提升阅读体验

  \caption{Overview of standard PDE benchmarks. Details on geometry type, spatial dimension, discretization size, dataset splits, and input features.}
  \vspace{2mm}
  \label{tab:dataset_overview}

  % 如果发现表格稍微超宽，可以取消下面 \resizebox 的注释
  % \resizebox{\textwidth}{!}{
  \begin{tabular}{l l c c c c l}
    \toprule
    \multirow{2}{*}{\textbf{Test Case}} & \multirow{2}{*}{\textbf{Geometry}} & \multirow{2}{*}{\textbf{Dim.}} & \multirow{2}{*}{\textbf{Resolution}} & \multicolumn{2}{c}{\textbf{Dataset Split}} & \multirow{2}{*}{\textbf{Input Type}} \\
    \cmidrule(lr){5-6}
    & & & & Train & Test & \\
    \midrule

    Airfoil \cite{li2023fourier}           & Structured Mesh & 2   & $221 \times 51$  & 1000 & 200 & Boundary Shape \\
    Pipe \cite{li2023fourier}              & Structured Mesh & 2   & $129 \times 129$ & 1000 & 200 & Domain Shape \\
    Plasticity \cite{li2023fourier}        & Structured Mesh & 2+1 & $101 \times 31$  & 900  & 80  & External Force \\
    Navier-Stokes \cite{li2020fourier} & Regular Grid    & 2+1 & $64 \times 64$   & 1000 & 200 & Previous Vorticity \\
    Darcy \cite{li2020fourier}         & Regular Grid    & 2   & $85 \times 85$   & 1000 & 200 & Coefficients \\
    Elasticity \cite{li2023fourier}    & Point Cloud     & 2   & 972         & 1000 & 200 & Domain Shape \\

    \bottomrule
  \end{tabular}
  % }
\end{table}
This benchmark includes a diverse range of physical systems (e.g., fluid dynamics and solid mechanics), geometry types (regular grids, structured meshes, and point clouds), and input functions.
The underlying physical equations and dataset details are introduced as in Appendix ~\ref{eq1}.

The quantitative results reported in Table~\ref{tab:main-benchmarks} demonstrate that CANO consistently outperforms the considered baselines across all six standard PDE benchmarks.
\begin{table}[!t]
  \centering
  % \small % 1. 设定固定字号 (和您之前的表格保持一致)
  \setlength{\tabcolsep}{4pt}
  \renewcommand{\arraystretch}{1.15} % 3. 设定行高，保持呼吸感

\caption{Performance comparison of neural operators on standard benchmarks (CANO vs. baselines). All values represent the relative $L^2$ error.
The best results are highlighted in \textbf{bold}, and the second-best are \underline{underlined}.
A slash (``/'') denotes benchmarks where the baseline is not applicable.
\textit{Models marked with $^*$ are reimplemented by us for fair comparison, while other results are taken from the original papers or the Transolver study.}
}
\vspace{1mm}
  \label{tab:main-benchmarks}

  \begin{tabular}{lcccccc}
    \toprule
    \multirow{3}{*}{\textbf{Model}} &
    \multicolumn{3}{c}{\textbf{Structured Mesh}} &
    \multicolumn{2}{c}{\textbf{Regular Grid}} &
    \multicolumn{1}{c}{\textbf{Point Cloud}} \\
    \cmidrule(lr){2-4}\cmidrule(lr){5-6}\cmidrule(lr){7-7}
      & Airfoil & Pipe & Plasticity & Navier-Stokes  & Darcy & Elasticity \\
      & $(\times 10^{-2})$ & $(\times 10^{-2})$ & $(\times 10^{-2})$ & $(\times 10^{-1})$ & $(\times 10^{-2})$ & $(\times 10^{-2})$ \\
    \midrule

    % Classic models
    \multicolumn{7}{l}{\bfseries\scshape Classic models} \\[0.3ex]
    FNO \cite{li2020fourier}        & / & / & / & $1.56$ & $1.08$ & / \\[0.2em]
    DeepONet \cite{lu2021learning}  & $3.85$ & $0.97$ & $1.35$ & $2.97$ & $5.88$ & $9.65$ \\[0.2em]
    U{-}FNO \cite{wen2022u}         & $2.69$ & $0.56$ & $0.39$ & $2.23$ & $1.83$ & $2.39$ \\[0.2em]
    Geo{-}FNO \cite{li2023fourier}  & $1.38$ & $0.67$ & $0.74$ & $1.56$ & $1.08$ & $2.29$ \\[0.2em]
    U{-}NO \cite{rahman2022u}       & $0.78$ & $1.00$ & $0.34$ & $1.71$ & $1.13$ & $2.58$ \\[0.2em]
    F{-}FNO \cite{tran2021factorized}& $0.78$ & $0.70$ & $0.47$ & $2.32$ & $0.77$ & $2.63$ \\[0.2em]
    LSM \cite{wu2023solving}        & $0.59$ & $0.50$ & $0.25$ & $1.54$ & $0.65$ & $2.18$ \\

    \midrule

    % Transformer models
    \multicolumn{7}{l}{\bfseries\scshape Transformer-based models} \\[0.3ex]
    Galerkin$^*$ \cite{cao2021choose}     & $0.64$ & $0.44$ & $0.17$ & $1.04$ & $1.97$ & $0.84$ \\[0.2em]
    HT{-}Net \cite{liu2024mitigating}     & $0.65$ & $0.59$ & $3.33$ & $1.85$ & $0.79$ & / \\[0.2em]
    OFormer \cite{li2022transformer}      & $1.83$ & $1.68$ & $0.17$ & $1.71$ & $1.24$ & $1.83$ \\[0.2em]
    GNOT \cite{hao2023gnot}               & $0.76$ & $0.47$ & $3.36$ & $1.38$ & $1.05$ & $0.86$ \\[0.2em]
    FactFormer \cite{li2023scalable}      & $0.71$ & $0.60$ & $3.12$ & $1.21$ & $1.09$ & / \\[0.2em]
    ONO \cite{xiao2023improved}           & $0.61$ & $0.52$ & $0.48$ & $1.19$ & $0.76$ & $1.18$ \\
    Transolver$^*$ \cite{wu2024transolver}& $\underline{0.51}$ & $\underline{0.42}$ & $\underline{0.12}$ & $\underline{0.94}$ & $\underline{0.52}$ & $\underline{0.56}$ \\

    \midrule
    \textbf{CANO (ours)} &
    $\mathbf{0.43}$ & $\mathbf{0.33}$ & $\mathbf{0.07}$ & $\mathbf{0.66}$ & $\mathbf{0.41}$ & $\mathbf{0.46}$ \\
    \midrule
    \textbf{Improvement} & 15.69\% & 21.43\% & 41.67\% & 29.79\% & 21.15\% & 17.86\% \\
    \bottomrule
  \end{tabular}
\end{table}
Compared to Transolver, CANO consistently achieves lower relative $L^2$ errors across all cases. The largest improvements occur on the (2+1)-D Plasticity dataset, where the error drops by 41.67\%, and on turbulent Navier--Stokes, with a 29.79\% reduction. These gains confirm that preserving full-resolution keys and values helps resolve both large solid deformations and turbulent flow dynamics.

\begin{figure}
  \centering
 \includegraphics[width=1.0\textwidth]{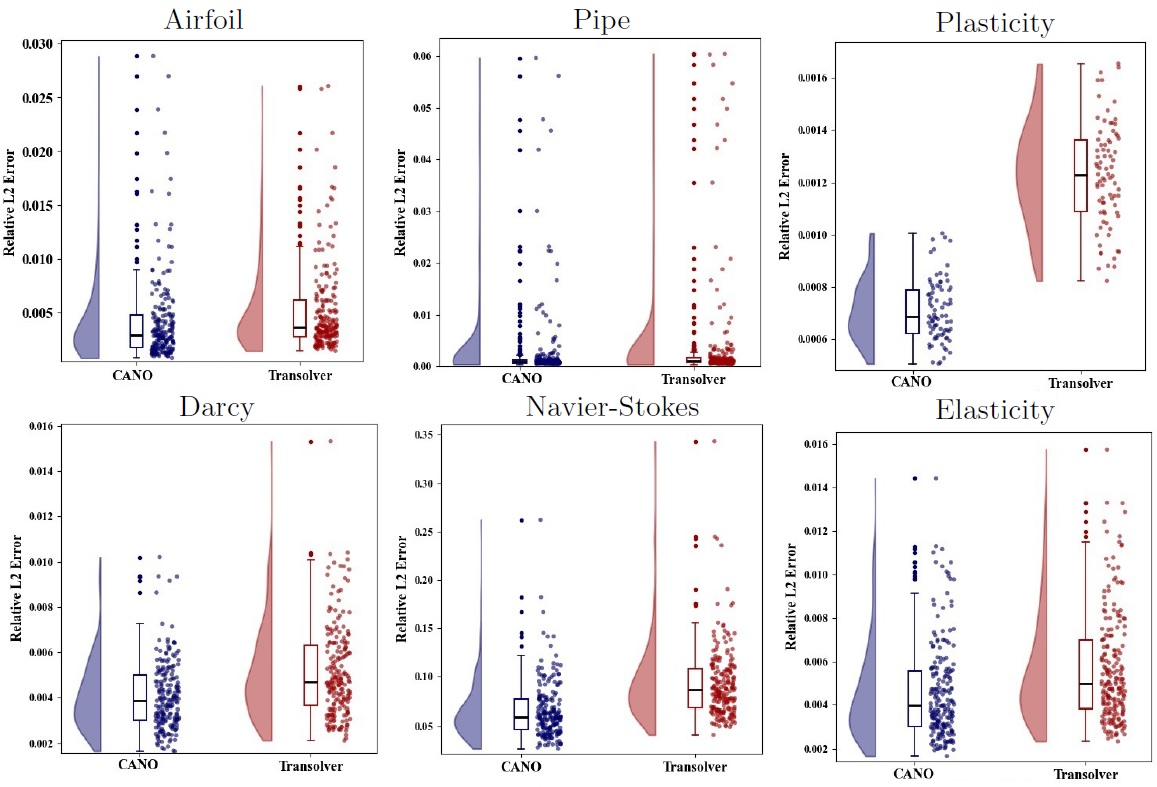}
\vspace{-0.2in}\caption{Performance and distribution comparison: Relative $L^2$ error of CANO and Transolver on six standard PDE datasets.}
\label{rain1}
\end{figure}

Figure~\ref{rain1} provides a complementary view of the comparison between CANO and Transolver by showing the distribution of the relative $L^2$ error across test samples.
CANO exhibits lower median errors on all six datasets, consistently with the results reported in Table~\ref{tab:main-benchmarks}. The difference is particularly marked for Plasticity, where the two distributions are clearly separated. In addition, CANO generally shows narrower interquartile ranges and more compact error distributions, indicating improved consistency across different test cases.

\begin{figure}[!t]
    \centering
\includegraphics[width=1.0\textwidth]{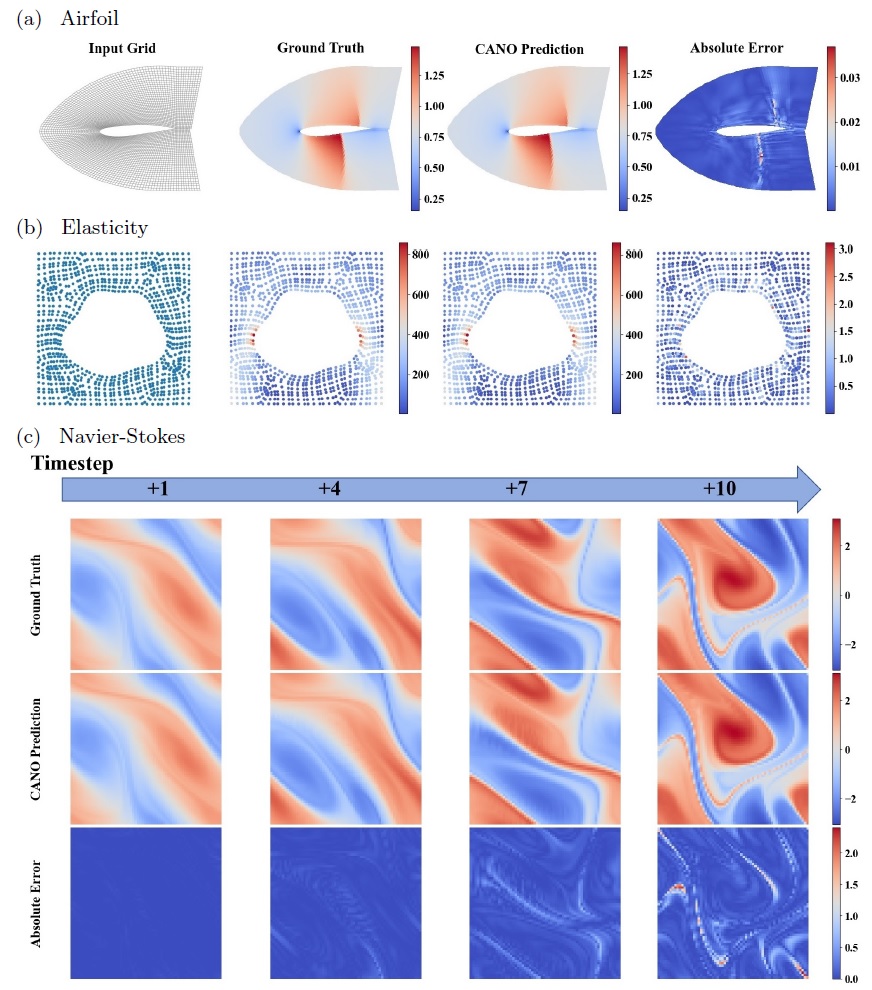}
\caption{CANO predictions on representative PDE benchmarks. Steady-state physical fields and absolute errors for \textbf{(a)} Airfoil (structured mesh) and \textbf{(b)} Elasticity (point cloud). CANO accurately captures sharp gradients near complex boundaries. \textbf{(c)} Long-term temporal rollout predictions (from timestep +1 to +10) on the Navier-Stokes dataset. CANO effectively preserves intricate high-frequency eddies and structural integrity over time without severe error accumulation.}
  \label{Vis1}
\end{figure}

Figure~\ref{Vis1} shows representative CANO predictions and absolute errors for three problems. On both the structured Airfoil mesh and the point-cloud Elasticity dataset, CANO captures the key spatial features with lower prediction errors. The Navier-Stokes example further shows that the model preserves the main flow structures over the autoregressive rollout up to timestep $+10$, with limited error accumulation.

\begin{figure}
    \centering
\includegraphics[width=1.0\textwidth]{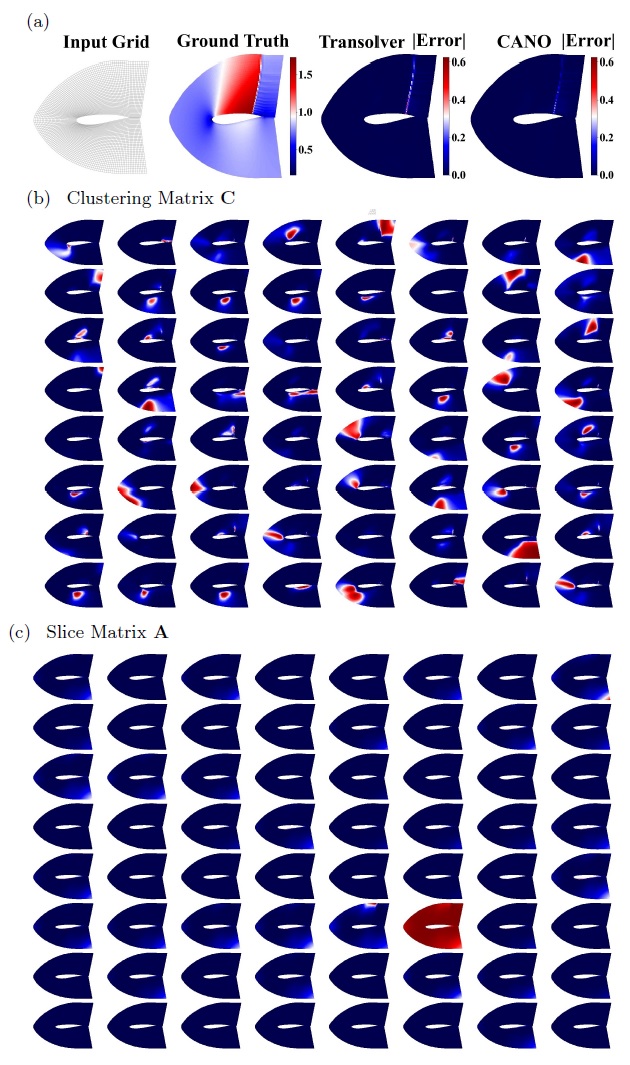}
\caption{Visualization of the  spatial assignments extracted from the final layer of the first head on the Airfoil dataset.
\textbf{(a)} The corresponding ground truth, predictions and absolute errors.
\textbf{(b)} The spatial activation maps of CANO's cluster matrix $C$, showing allocations around  discontinuities (e.g., shockwaves).
\textbf{(c)} The slice allocation matrix of Transolver $A$, displaying relatively smooth regions.}
  \label{fig:matrix_vis}
\end{figure}

To further examine the behavior of the proposed attention mechanism, Figure~\ref{fig:matrix_vis} compares the spatial organization learned by CANO and Transolver on the Airfoil benchmark. In particular, we visualize the latent assignment matrices extracted from the final layer of the first attention head: the cluster matrix $\mathbf{C}\in\mathbb{R}^{N\times K_c}$ for CANO and the slice matrix $\mathbf{A}\in\mathbb{R}^{N\times M}$ for Transolver. In CANO, the assignment matrix $\mathbf{C}$ determines how query matrix $Q$ are mapped to  query clusters $Q_c$, while in Transolver, the slice matrix $A$ determines  the probability that the original token $X$ belongs to the slice token $S$. As shown in Figure~\ref{fig:matrix_vis}(b), these cluster assignments are highly localized, with several clusters concentrating around regions with strong flow gradients---specifically near the shockwave. This confirms that CANO adapts its latent clusters to physical features rather than simply partitioning the domain by geometric proximity. In contrast, the slice assignments in Transolver (Figure~\ref{fig:matrix_vis}(c)) are  smoother, showing little localized adaptation around flow discontinuities. While both methods capture the overall flow field (Figure~\ref{fig:matrix_vis}(a)), CANO yields a significantly  lower  prediction errors.

For neural operators, a low global error (e.g., relative $L^2$ norm) does not necessarily imply physical fidelity in the predicted fields.
% Highly compressed efficient architectures often suffer from severe spectral bias \ac{CITEAAA}, acting as unintended low-pass filters that smooth out intricate small-scale eddies and sharp gradients.
To evaluate spectral accuracy across spatial scales, we compute the turbulent kinetic energy (TKE) spectrum of the predicted voticity fields.

To compute the one-dimensional (1D) kinetic energy spatial spectrum $E(k)$ from a two-dimensional (2D) predicted vorticity field $w(\mathbf{x})$ of the Navier-Stokes benchmark, we first obtain its Fourier coefficients in the wavenumber space:
\begin{equation}
    \hat{w}(\mathbf{k}) = \mathcal{F}\{w(\mathbf{x})\},
\end{equation}
where $\mathbf{k} = (k_x, k_y)$ is the wavenumber vector and $k = \vert\mathbf{k}\vert = \sqrt{k_x^2 + k_y^2}$. By applying the incompressibility condition, the 2D spectral kinetic energy density is derived by scaling the vorticity power spectrum by the inverse square of the wavenumber (excluding the zero-frequency mean flow where $\vert\mathbf{k}\vert = 0$):
\begin{equation}
    E(\mathbf{k}) = \frac{1}{2} \frac{\vert\hat{w}(\mathbf{k})\vert^2}{\vert\mathbf{k}\vert^2}.
\end{equation}
Finally, the 1D radial energy spectrum is calculated by isotropically summing the discrete 2D energy density over concentric annuli of radius $k$ and bin width $\Delta k$:
\begin{equation}\label{TKE}
   \mathbf{E(k)} = \sum_{k \le \vert\mathbf{k}\vert < k+\Delta k} E(\mathbf{k}),
\end{equation}
which effectively quantifies the distribution of turbulent kinetic energy across varying spatial scales.

\begin{figure}
  \centering
\includegraphics[width=1.0\textwidth]{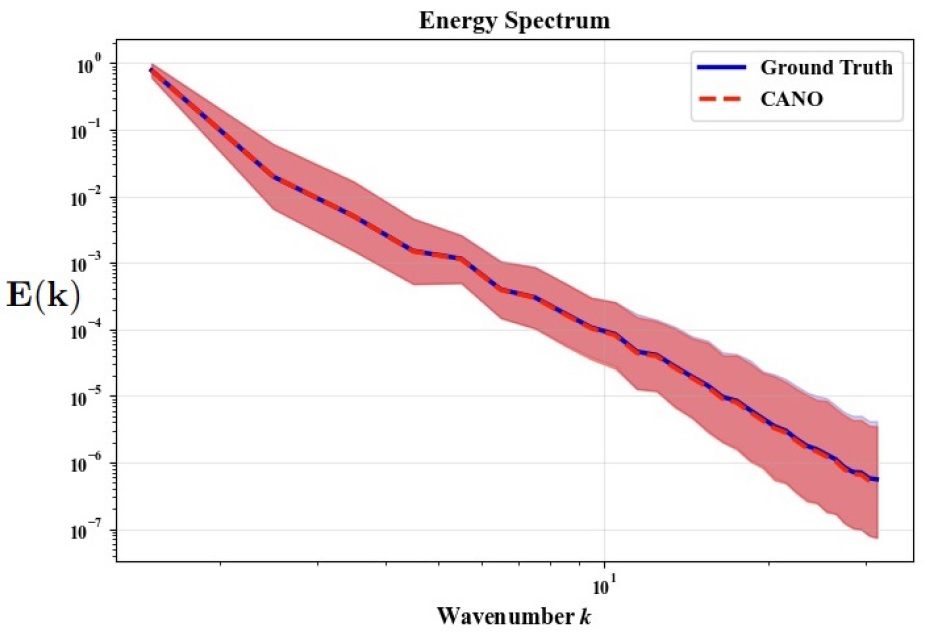}
%\vspace{0.22in}
\caption{Comparison of kinetic energy spectra (\ref{TKE}) between CANO predictions and ground truth for the Navier-Stokes benchmark. Solid lines show the mean spectrum over 200 test samples; shaded regions indicate the mean $\pm$ 3 standard deviation.}
\label{Spectral}
\end{figure}

Based on this formulation, Figure \ref{Spectral} visualizes the energy spectra $E(k)$ of the ground truth and the CANO predicted fields, for the Navier-Stokes benchmarch, averaged over 200 test samples.
The spectral distribution predicted by CANO (dashed red line) matches well with the reference solution (solid blue line) across the entire wavenumber domain $k$, with the shaded regions (mean $\pm$ 3 standard deviation) largely overlapping. Even at small scales where the energy drops significantly, the spectrum predicted by CANO aligns well with the reference data.

\begin{table}[!t]
  \centering
  \renewcommand{\arraystretch}{1.3}
  \setlength{\tabcolsep}{18pt}
  \caption{Comparison of model parameters between Transolver and CANO across different standard cases.}
  \vspace{2mm}
  \label{tab:model_parameters}
  \begin{tabular}{lcc}
    \toprule
    \multirow{2}{*}{\textbf{Benchmark}} & \multicolumn{2}{c}{\textbf{Number of Parameters}} \\
    \cmidrule(lr){2-3}
    & \textbf{Transolver} & \textbf{CANO (Ours)} \\
    \midrule
    Airfoil       & 3.07M & 2.44M \\
    Pipe          & 3.07M & 2.49M \\
    Plasticity    & 3.11M & 2.51M \\
    Navier-Stokes   & 12.28M & 8.12M \\
    Darcy         & 3.09M & 2.42M \\
    Elasticity    & 0.98M & 1.33M \\
    \bottomrule
  \end{tabular}
\end{table}
To evaluate model efficiency, we compare the parameter counts of CANO and Transolver across all six examples (Table~\ref{tab:model_parameters}). Overall, CANO achieves higher accuracy while maintaining a more compact model size. In five out of the six benchmarks, CANO requires fewer parameters than Transolver. This difference is most noticeable on Navier--Stokes, where CANO reduces the parameter count from 12.28M to 8.12M (a 33.9\% reduction). Although CANO uses slightly more parameters on the Elasticity dataset, the overall results show that it is more accurate while using fewer  parameters.

\subsection{Irregular domain benchmarks}

We next evaluate CANO on irregular geometries, testing its ability to handle complex physical domains. Table \ref{tab:irregular_dataset_overview} summarizes the three irregular domain PDE examples \cite{chen2023} utilized in this section: Irregular Darcy, Pipe Turbulence, and Composite.
\begin{table}[htbp]
  \centering
  \small % 保持字号缩小，让表格显得精致
  \renewcommand{\arraystretch}{1.2} % 增加行高，提升阅读体验

  \caption{Overview of irregular domain PDE benchmarks. Details on geometry type, spatial dimension, node discretization, dataset splits, and input features.}
  \vspace{2mm}
  \label{tab:irregular_dataset_overview}

  % 如果发现表格稍微超宽，可以取消下面 \resizebox 的注释
  % \resizebox{\textwidth}{!}{
  \begin{tabular}{l l c c c c l}
    \toprule
    \multirow{2}{*}{\textbf{Test Case}} & \multirow{2}{*}{\textbf{Geometry}} & \multirow{2}{*}{\textbf{Dim.}} & \multirow{2}{*}{\textbf{Nodes ($N$)}} & \multicolumn{2}{c}{\textbf{Dataset Split}} & \multirow{2}{*}{\textbf{Input Type}} \\
    \cmidrule(lr){5-6}
    & & & & Train & Test & \\
    \midrule

    Irregular Darcy \cite{chen2023} & Point Cloud & 2 & 2290 & 1000 & 200 & Coefficients \\
    Pipe Turbulence \cite{chen2023} & Point Cloud & 2 & 2673 & 300  & 100 & Previous State \\
    Composite \cite{chen2023}       & Point Cloud & 3 & 8232 & 400  & 100 & Temperature \\

    \bottomrule
  \end{tabular}
  % }
\end{table}
Unlike the previous benchmarks that mostly use regular grids or structured meshes, all geometries in this task are represented entirely as 2D or 3D point clouds, posing significant challenges. We investigate two 2D fluid problems: the Irregular Darcy flow (2,290 nodes) driven by varying the permeability coefficient field, and the Pipe Turbulence (2,673 nodes) requiring state predictions based on previous one. Furthermore, we test our evaluation to a challenging 3D solid mechanics task using the Composite material benchmark (8,232 nodes) driven by temperature profiles. These benchmarks are useful to validate the model's capability to process  irregular geometries.  The underlying physical equations and dataset details are introduced as in Appendix ~\ref{eq2}.

A quantitative comparison with other baselines across these three cases is presented in Table \ref{tab:new-benchmarks}.
\begin{table}[!t]
  \centering
  \small % 建议用 small 代替默认大小
  \setlength{\tabcolsep}{8pt} % 减小列间距，让表格更紧凑
  \renewcommand{\arraystretch}{1.1} % 稍微收缩垂直间距

\caption{Performance comparison of neural operators on Irregular Darcy, Pipe Turbulence, and Composite benchmarks (CANO vs. baselines). All values represent the relative $L^2$ error $(\times 10^{-4})$.
The best results are highlighted in \textbf{bold}, and the second-best are \underline{underlined}.
}
  \vspace{2mm}
  \label{tab:new-benchmarks}

  \begin{tabular}{lccc}
    \toprule
    \textbf{Model} & \textbf{Irregular Darcy} & \textbf{Pipe Turbulence} & \textbf{Composite} \\
    \midrule
    GraphSAGE \cite{hamilton2017} & 6.73 & 23.60 & 20.90 \\[0.1em]
    DeepONet \cite{lu2021learning} & 1.36 & 9.36 & 1.88 \\[0.1em]
    POD-DeepONet \cite{lu2022pod} & 1.30 & 2.59 & 1.44 \\[0.1em]
    NORM \cite{chen2023} & 1.05 & 1.01 & 1.00 \\[0.1em]
    HPM \cite{yue2024} & \underline{0.74} & \underline{0.83} & \underline{0.93} \\[0.1em]
    Transolver$^*$ \cite{wu2024transolver} & 0.89 & 1.01 & 1.61 \\[0.1em]
    \midrule
    \textbf{CANO (ours)} & \textbf{0.71} & \textbf{0.55} & \textbf{0.89} \\[0.1em]
    \textbf{Improvement} & 4.05\% & 33.73\% & 4.30\% \\
    \bottomrule
  \end{tabular}
\end{table}
CANO consistently achieves the lowest relative $L^2$ error, outperforming baseline models ranging from foundational networks (GraphSAGE, DeepONet) to recent operators (NORM, HPM, and Transolver). Specifically, on the Pipe Turbulence dataset, CANO yields a 33.73\% error reduction compared to the second-best baseline. Furthermore, while Transolver performs competitively on standard benchmarks, its error increases on these unstructured point clouds, particularly in the 3D Composite task. This observation suggests that the slicing mechanism struggles to effectively compress highly unstructured and non-uniform spatial information. In contrast, CANO's cross-attention mechanism keeps keys and values at their full spatial resolution, maintaining stable predictive accuracy across different geometric discretizations.

\begin{figure}
    \centering
\includegraphics[width=0.9\textwidth]{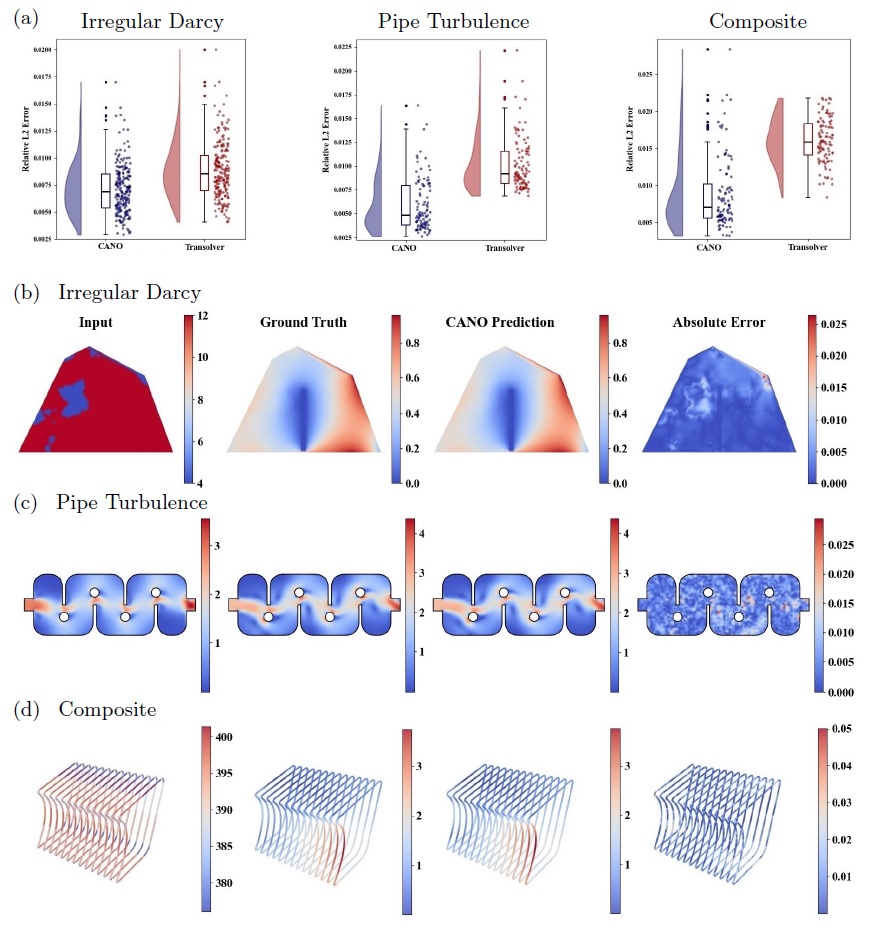}
\vspace{-0.1in}
\caption{Performance and qualitative distribution on irregular domains. \textbf{(a)} Raincloud plots comparing the relative $L^2$ error distributions of CANO and Transolver.  \textbf{(b)-(d)} Representative visual predictions of CANO for the \textbf{(b)} Irregular Darcy, \textbf{(c)} Pipe Turbulence, and \textbf{(d)} Composite benchmarks.}
\label{Vis2}
\end{figure}

Figure \ref{Vis2} presents both the error distributions and the spatial distributions of the predictions.
Panel (a) displays raincloud plots comparing CANO and Transolver across all three datasets. CANO shifts the overall error distribution downward and suppresses extreme outliers in Irregular Darcy and Pipe Turbulence datasets. While some outlier errors occur in Composite dataset, the overall average error of CANO is still lower than that of Transolver.
Panels~(b--d) show qualitative predictions for the 2D irregular Darcy flow, Pipe Turbulence, and 3D Composite material, respectively. The predicted fields match the ground truth closely, and the corresponding absolute error maps confirm that CANO maintains low errors, which highlights its effectiveness on irregular physical geometries.

\subsection{Long-time rollout benchmarks}

In practical physical simulations, neural operators are frequently required to extrapolate system dynamics far beyond the short temporal horizons observed during training. The challenge in such cases is to avoid severe error accumulation, numerical dissipation, and the eventual collapse of physical structures over time.

The autoregressive stability and temporal generalization capabilities of our proposed model are evaluated through the three time-dependent datasets shown in Table \ref{tab:rw_benchmarks}: Navier-Stokes~\cite{li2020fourier}, Inductively Coupled Plasma (ICP) and Heat Transfer~\cite{liu2025}. Specifically, the Navier-Stokes dataset models the evolution of chaotic fluid vorticity ($\omega$) on a regular $64 \times 64$ grid. In contrast, the ICP Plasma and Heat Transfer datasets involve highly nonlinear multi-physics phenomena defined on  point clouds, predicting complex state variables such as electron density, electron temperature, and velocity fields.
\begin{table}
  \centering
  \small
  \renewcommand{\arraystretch}{1.2}
  \caption{Overview of 2D long-time rollout benchmarks. Details on geometry type, node discretization, training/test window, dataset splits, and input features. Models are trained on short temporal windows ($T_{\text{out}}=5$) but evaluated on significantly longer autoregressive rollouts (up to 15 or 50 steps) to test temporal stability.}
  \vspace{2mm}
  \label{tab:rw_benchmarks}

  \small
  \setlength{\tabcolsep}{1pt}
  \begin{tabular}{l l c c c c c c c l}
    \toprule
    \multirow{2}{*}{\textbf{Test Case}} & \multirow{2}{*}{\textbf{Geometry}} & \multirow{2}{*}{\textbf{Nodes}} &
    \multicolumn{2}{c}{\textbf{Train Window}} & \multicolumn{2}{c}{\textbf{Test Window ($T_{\text{out}}$)}} & \multicolumn{2}{c}{\textbf{Data Split}} & \multirow{2}{*}{\textbf{Input Type}} \\
    \cmidrule(lr){4-5} \cmidrule(lr){6-7} \cmidrule(lr){8-9}
    & & & $T_{\text{in}}$ & $T_{\text{out}}$ & Short & Long & Train & Test & \\
    \midrule

    Navier-Stokes~\cite{li2020fourier} & Regular Grid & $64 \times 64$ & 5  & 5  & 5 & 15 & 1000 & 200 & $\omega$ \\
    ICP Plasma~\cite{liu2025}    & Point Cloud  & 3424         & 5  & 5  & 5 & 50 & 499  & 150 & $n_e,T_e,v,T$ \\
    Heat Transfer~\cite{liu2025} & Point Cloud  & 4887 (Avg.)  & 5  & 5  & 5 & 50 & 740  & 132 & $u,v$ \\
    \bottomrule
  \end{tabular}
\end{table}

We use random-window training to enhance the model's temporal stability across rollouts. During the training phase, all models can only see the historical  of $T_{\text{in}}=5$ steps, learning to predict the subsequent $T_{\text{out}}=5$ steps. However, during the inference phase, the models are evaluated on two distinct scenarios: a short rollout matching the training horizon ($T_{\text{out}}=5$), and a significantly long autoregressive rollout, extrapolating up to $T_{\text{out}}=15$ steps for the  Navier-Stokes dataset, and up to $T_{\text{out}}=50$ steps for both the ICP Plasma and Heat Transfer datasets.

Table~\ref{tab:temporal_evolution} shows that CANO consistently outperforms Transolver throughout the autoregressive rollout. Although the relative $L^2$ error increases with the prediction horizon for both models, CANO maintains lower errors across all datasets and time horizons, indicating better long-term predictive stability.
\begin{table}
  \centering
  \small
  \renewcommand{\arraystretch}{1.2}
  \caption{Long-term predictive performance across different datasets. We evaluate the relative $L^2$ error $(\times 10^{-1})$ at different future time steps (e.g.,  +5,  +15,  +50). The best results are highlighted in \textbf{bold}.}
  \vspace{1mm}
  \label{tab:temporal_evolution}
  \begin{tabular}{l cc cc cc}
    \toprule
    \multirow{2}{*}{\textbf{Model}} & \multicolumn{2}{c}{\textbf{Navier-Stokes}} & \multicolumn{2}{c}{\textbf{ICP Plasma}} & \multicolumn{2}{c}{\textbf{Heat Transfer}} \\
    \cmidrule(lr){2-3} \cmidrule(lr){4-5} \cmidrule(lr){6-7}
    & +5 & +15 & +5 & +50 & +5 & +50 \\
    \midrule
    Transolver       & 1.21 & 3.22 & 0.29 & 2.11 & 0.70 & 4.87 \\
    \textbf{CANO}    & \textbf{0.59} & \textbf{2.57} & \textbf{0.23} & \textbf{1.61} & \textbf{0.58} & \textbf{4.19} \\
    \bottomrule
  \end{tabular}
\end{table}

\begin{figure}
    \centering
\includegraphics[width=1.0\textwidth]{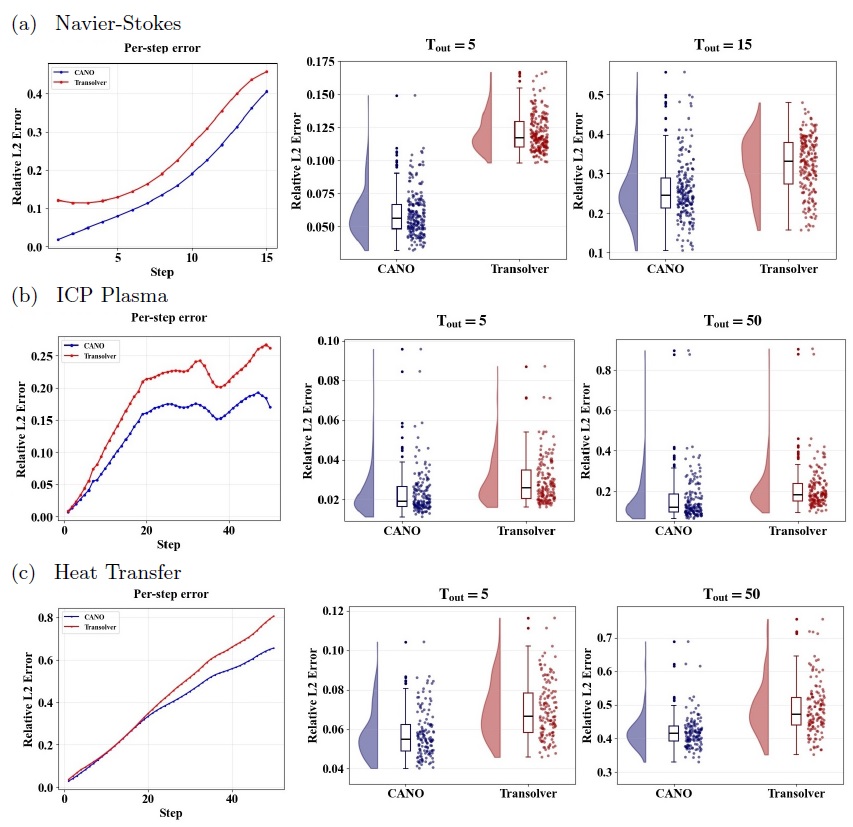}
\vspace{-0.1in}
\caption{Temporal error evolution and distribution analysis for long time rollouts. Panels \textbf{(a)}, \textbf{(b)}, and \textbf{(c)} correspond to the Navier-Stokes, ICP Plasma, and Heat Transfer datasets, respectively. \textbf{Left:} Per step relative $L^2$ error accumulation curves, showing CANO's slower error growth rate. \textbf{Middle and Right:} Raincloud plots comparing the error distributions at the short ($T_{\text{out}}=5$) and long ($T_{\text{out}}=15$ or $50$) rollout stages.}
  \label{fig:long_rollout_analysis}
\end{figure}

Figure \ref{fig:long_rollout_analysis} presents the corresponding error evolution and statistical distributions. The left column displays the relative $L^2$ error accumulation per step, where the errors  for both models increases as the simulation advances. The middle and right columns show the error distributions at the short ($T_{\text{out}}=5$) and long-time horizons. While the error distributions for both models are clustered at $T_{\text{out}}=5$, the spread of the distribution increases for both models at the long-time horizons. In these extended rollouts, Transolver shows a more pronounced upper tail in the error distribution than CANO.

\begin{figure}
  \centering
  \includegraphics[width=1.0\textwidth]{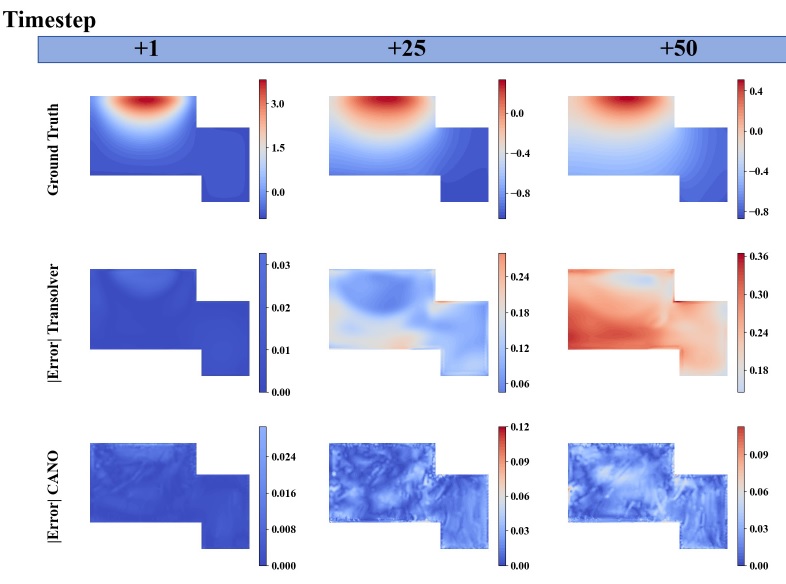}
  \vspace{-0.1in}
  \caption{Temporal error evolution on the ICP Plasma dataset. The visualization displays the absolute error of the predicted electron temperature $T_e$ for Transolver and CANO at timesteps +1, +25, and +50 compared to the ground truth.}
  \label{Vis3}
\end{figure}

\begin{figure}
  \centering
  \includegraphics[width=1.02\textwidth]{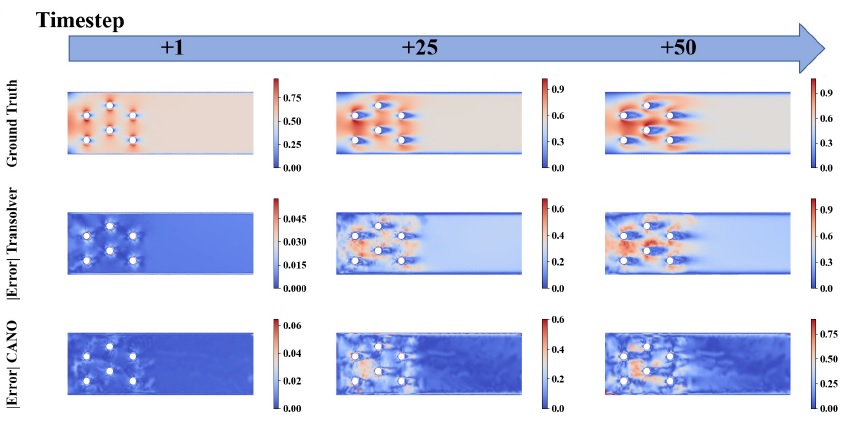}
  \vspace{-0.1in}
  \caption{Temporal error evolution on the Heat Transfer dataset. The visualization displays the absolute error of the predicted velocity magnitude ($\sqrt{u^2 + v^2}$) for Transolver and CANO at timesteps +1, +25, and +50 compared to the ground truth.}
  \label{Vis4}
\end{figure}

Figures~\ref{Vis3} and~\ref{Vis4} show the spatial distribution of the absolute error during the autoregressive rollout for the ICP Plasma dataset, considering the electron temperature $T_e$, and the Heat Transfer dataset, considering the velocity magnitude $\sqrt{u^2+v^2}$. At the first several prediction steps, CANO and Transolver exhibit comparable error distributions. As the rollout progresses, however, the errors produced by Transolver become increasingly pronounced across the domain, while CANO maintains consistently lower error levels over time.

\begin{table}[!t]
  \centering
  \renewcommand{\arraystretch}{1.2}
  \caption{Relative $L^2$ error comparison across different training sample sizes. All values for Darcy Flow are scaled by $10^{-2}$, and Navier-Stokes by $10^{-1}$. Best results are highlighted in \textbf{bold}.}
  \vspace{2mm}
  \label{tab:sample_efficiency}

  \begin{tabular}{lccc}
    \toprule
    \textbf{Problem} & \textbf{Training Number} & \textbf{Transolver} & \textbf{CANO (ours)} \\
    \midrule

    \multirow{5}{*}{\textbf{Darcy Flow}}
     & 200  & $2.17$ & $\mathbf{1.80}$ \\
     & 400  & $0.98$ & $\mathbf{0.72}$ \\
     & 600  & $0.71$ & $\mathbf{0.54}$ \\
     & 800  & $0.67$ & $\mathbf{0.46}$ \\
     & 1000 & $0.52$ & $\mathbf{0.41}$ \\

    \midrule

    \multirow{5}{*}{\textbf{Navier-Stokes}}
     & 200  & $3.76$ & $\mathbf{3.09}$ \\
     & 400  & $3.14$ & $\mathbf{1.96}$ \\
     & 600  & $2.87$ & $\mathbf{1.01}$ \\
     & 800  & $2.49$ & $\mathbf{0.90}$ \\
     & 1000 & $0.94$ & $\mathbf{0.66}$ \\

    \bottomrule
  \end{tabular}
\end{table}

\subsection{Model Analysis}

In this section, we analyze how CANO scales with training sample size and evaluate its sensitivity to the number of  clusters ($K_c$).

\paragraph{Training Sample Efficiency}
Table \ref{tab:sample_efficiency} evaluates the relative $L^2$ error on the Darcy Flow and Navier-Stokes datasets when the number of training samples is scaled from 200 to 1000. Both evaluated models exhibit a monotonic decrease in error as the data scale expands. Under identically restricted data regimes, CANO consistently maintains lower error bounds compared to the Transolver baseline. Notably, in the most data-scarce scenario (200 samples), CANO limits the error to $1.80 \times 10^{-2}$ on Darcy Flow and $3.09 \times 10^{-1}$ on Navier-Stokes, yielding a better performance than that of Transolver ($2.17 \times 10^{-2}$ and $3.76 \times 10^{-1}$, respectively).
\begin{table}
  \centering
  \small
  \setlength{\tabcolsep}{4pt}
  \renewcommand{\arraystretch}{1.2} % Increase row height slightly for better readability
  \caption{Relative $L^2$ errors of CANO with different cluster numbers ($K_c$) across various PDE datasets. The best results for each dataset are highlighted in \textbf{bold}.}
  \vspace{2mm}
  \begin{tabular}{ccccccc}
    \toprule
    \textbf{Cluster Numeber ($K_c$)} & \textbf{Darcy} & \textbf{Navier-Stokes } & \textbf{Pipe} & \textbf{Airfoil} & \textbf{Elasticity} & \textbf{Plasticity} \\
    % \textbf{Num ($K_c$)} & ($\times 10^{-3}$) & ($\times 10^{-2}$) & ($\times 10^{-3}$) & ($\times 10^{-3}$) & ($\times 10^{-3}$) & ($\times 10^{-3}$) \\
    \midrule

    8   & $5.38$ & $9.05$ & $3.83$ & $4.96$ & $5.08$ & $0.88$ \\
    16  & $4.83$ & $7.67$ & $\mathbf{3.32}$ & $4.77$ & $4.78$ & $0.78$ \\
    32  & $4.73$ & $6.58$ & $3.55$ & $\mathbf{4.29}$ & $4.72$ & $0.81$ \\
    64  & $4.36$ & $6.36$ & $3.48$ & $4.88$ & $4.81$ & $\mathbf{0.71}$ \\
    128 & $\mathbf{4.08}$ & $\mathbf{5.60}$ & $3.82$ & $4.38$ & $\mathbf{4.62}$ & $0.77$ \\

    \bottomrule
  \end{tabular}
  \label{tab:cluster_ablation}
\end{table}

\paragraph{Sensitivity to Cluster Number}
Table~\ref{tab:cluster_ablation} shows the effect of the cluster number $K_c$ across six datasets, with $K_c$ varied from 8 to 128. The optimal choice of $K_c$ depends on the specific physical system. For complex or turbulent flows (e.g., Navier--Stokes and Darcy Flow), the error systematically decreases with larger $K_c$, reaching the lowest value at $K_c=128$. In contrast, for systems governed by localized dynamics (Pipe, Airfoil, and Plasticity), the lowest relative $L^2$ errors occur at moderate configurations ($K_c=16$, $32$, and $64$, respectively). This indicates that aligning $K_c$ with the spatial complexity of the PDE helps prevent representational redundancy.
% ... existing code ...

\section{Conclusions and Discussion}
\label{sec:conclusions}
In this work, we introduced the Cluster Attention Neural Operator (CANO) to address the information loss caused by latent space compression. CANO relies on an asymmetric cross-attention mechanism that clusters the query space while keeping keys and  at full spatial resolution. This formulation allows the model to preserve fine-scale information while maintaining linear $\mathcal{O}(N)$ computational complexity. Experiments across 12 PDE benchmarks, including regular and irregular geometries as well as long-horizon autoregressive prediction, show that CANO consistently improves predictive accuracy over the considered state-of-the-art baselines.

Several directions remain open for future work. In the present work, the number of clusters $K_c$ is fixed as a priori. An adaptive strategy that adjusts the cluster number according to  solution complexity could provide greater performance. In addition, although CANO retains linear complexity, optimized implementations of the clustering and cross-attention operations will be important for extending the method to large-scale three-dimensional applications~\cite{luo2025transolver,nabian2024xmeshgraphnet,bleeker2025neuralcfd}.

The results suggest that asymmetric cluster-based attention provides an effective alternative to symmetric latent compression for neural operator learning, particularly when fine-scale spatial information and long-term predictive stability are important. Overall, CANO offers a framework for scientific machine learning, enabling accurate and efficient modeling of complex  physical systems.

\vspace{0.2in} \noindent {\bf Declaration of Competing Interest}

\vspace{0.05in} The authors declare that they have no known competing financial interests or personal relationships that could have appeared to
influence the work reported in this paper.

\vspace{0.2in} \noindent {\bf Data availability}

\vspace{0.05in} The data that support the findings of this study are available
within the article.

\vspace{0.2in}
\noindent {\bf Acknowledgments}

\vspace{0.05in}
This work was partially supported by the National Key R\&D Program of China (No. 2024YFA1013101), the National Natural Science Foundation of China (No. 12471242), and the China Scholarship Council.

\appendix
\section{Theoretical Connection to Linear Attention}
\label{link}

To motivate CANO mathematically, we analyze existing efficient neural operators through the unified framework of generalized linear attention~\cite{2026linearno}. To decouple the quadratic Softmax computation, the generalized linear attention framework employs independent feature mappings $\phi(\cdot)$ and $\psi(\cdot)$, and further incorporates an intermediate operation $G$ to process the projected latent representations. The attention mechanism can thus be reformulated as~\cite{2026linearno}:
\begin{equation}\label{gla}
    \mathbf{Y} = \phi(\mathbf{Q}) \circ G \circ \big( \psi(\mathbf{K})^\top \mathbf{V} \big).
\end{equation}

{\it Revisiting Transolver from a Generalized  Linear Attention Perspective}. Recent analysis reveals that the Physics-Attention mechanism in Transolver \cite{wu2024transolver} can be mathematically unified under this generalized linear attention paradigm. Specifically, Transolver compresses physical nodes into $M$ latent slices using a slice-weight matrix, applies an intermediate self-attention operation $G(\cdot)$ within this slice space, and subsequently deslices the representations back to the physical domain. In the context of the generalized formulation Eq.~(\ref{gla}), Transolver's feature mappings are constructed as:
\begin{align}
\phi_{\mathrm{Transolver}}(\mathbf{Q}) &= \mathrm{Softmax}(\mathbf{X}\mathbf{A}) \in \mathbb{R}^{N \times M}, \\
\psi_{\mathrm{Transolver}}(\mathbf{K})^\top &= \mathrm{Norm}\big(\mathbf{X}\mathbf{A}\big)^\top \in \mathbb{R}^{M \times N},
\end{align}
where $\mathbf{A} \in \mathbb{R}^{d \times M}$ is the learnable weight matrix for slicing.

Notably, $\phi_{\mathrm{Transolver}}$ and $\psi_{\mathrm{Transolver}}$ are derived from the exact same parameter $\mathbf{A}$, differing only in their normalization strategies.
 This architectural choice introduces two structural limitations. First, the forced parameter-sharing between the slicing ($\psi$) and deslicing ($\phi$) mappings constrains the model's performance (see Figure \ref{fig:matrix_vis} ). Second, because the intermediate interaction $G(\cdot)$ operates strictly within the highly compressed slice space, the model's ability to capture fine-grained  dynamics is difficult by the information loss  during the  geometric compression.

{\it CANO as Dynamic Generalized Linear Attention}. Our proposed CANO  resolves these limitations. Instead of entirely decoupling the queries and keys through independent linear projections in Eq.~(\ref{gla}), CANO dynamically clusters the queries while preserving the full-resolution keys, executing a  cross-attention mechanism.

Mathematically, by expanding our dispatching and aggregation equations, the full CANO computation is formulated as:
\begin{equation}
\mathbf{Y} = \mathbf{C} \, \mathrm{Softmax}\Big(\frac{(\mathbf{C}^\top \mathbf{Q}) \mathbf{K}^\top}{\sqrt{d}}\Big) \mathbf{V}.
\end{equation}
Then we can establish an exact equivalence to the generalized linear attention given by Eq.~(\ref{gla}) by defining the CANO feature mappings as:
\begin{align}
\phi_{\mathrm{CANO}}(\mathbf{Q}) &\triangleq \mathbf{Q}_c = \mathbf{C}^\top (\mathbf{X}\mathbf{W}_Q) \in \mathbb{R}^{K_c \times d}, \\
\psi_{\mathrm{CANO}}(\mathbf{K})^\top &\triangleq \mathrm{Softmax}\left(\frac{\mathbf{Q}_c (\mathbf{X}\mathbf{W}_K)^\top}{\sqrt{d}}\right) \in \mathbb{R}^{K_c \times N},
\end{align}
Finally the CANO output can be written as:
\begin{equation}
\mathbf{Y}_{\mathrm{CANO}} = \phi_{\mathrm{CANO}}(\mathbf{Q}) \Big(\psi_{\mathrm{CANO}}( \mathbf{K})^\top \mathbf{V} \Big).
\end{equation}

This highlights the main advantage of CANO's architecture. First, CANO is  asymmetric, effectively avoiding the structural issue of Transolver. The left mapping $\phi_{\mathrm{CANO}}(\mathbf{Q})$ acts as a dynamically computed cluster assignment derived strictly from the query space, while the right mapping $\psi_{\mathrm{CANO}}(\mathbf{K})$ processes the interaction between the clustered queries $\mathbf{Q}_c$ and the full resolution keys $\mathbf{K}$.   More importantly, unlike Transolver which compresses the entire input field, CANO preserves the keys and values in their original full resolution space.
% By formulating $\psi_{\mathrm{CANO}}$ as a joint function of clustered queries and full resolution keys, we embed a fully coupled Softmax operation between the clustered query centers and the uncompressed key tensors.

% Because $K_c \ll N$, CANO evaluates the term $\big( \psi_{\mathrm{CANO}}( \mathbf{K})^\top \mathbf{V} \big)$ with a  linear complexity of $\mathcal{O}(N K_c d)$. Consequently, CANO effectively realizes a dynamic generalized linear attention: it perfectly inherits the $\mathcal{O}(N)$ computational scalability of linear transformers, while fundamentally bypassing their fatal over-smoothing flaws by retaining the expressive, non-linear Softmax coupling within a computationally tractable subspace.

\section{Govering Equations}

In this part, we present the governing equations for the standard PDE benchmarks and the irregular domain benchmarks.

\subsection{Standard PDE Benchmarks}
\label{eq1}

\paragraph{Airfoil.} This dataset models the transonic flow over an airfoil. Given the low air viscosity in this scenario, the viscous term is negligible, and the system is governed by the Euler equations (representing mass, momentum, and energy conservation, respectively):
\begin{align}
    \frac{\partial \rho_f}{\partial t} + \nabla \cdot (\rho_f U) &= 0, \\
    \frac{\partial \rho_f U}{\partial t} + \nabla \cdot (\rho_f UU + pI) &= 0, \\
    \frac{\partial E}{\partial t} + \nabla \cdot ((E + p)U) &= 0,
\end{align}
where $\rho_f$ is the fluid density and $E$ is the total energy. The data is generated on a structured mesh with a resolution of $221 \times 51$. The model learns to map the locations of these mesh points to the Mach number.

\paragraph{Pipe.} This benchmark focuses on incompressible flow through a pipe. The governing equations are deduced from the Navier-Stokes relations for Newtonian fluids:
\begin{align}\label{eq:pipe}
    \nabla \cdot U &= 0, \\
    \frac{\partial U}{\partial t} + U \cdot \nabla U &= f - \frac{1}{\rho}\nabla p + \nu \nabla^2 U.
\end{align}
The dataset utilizes a structured mesh with a resolution of $129 \times 129$. We use the mesh structure as input to predict the horizontal fluid velocity within the pipe.

\paragraph{Plasticity.} Focusing on the plastic forging problem, this dataset simulates a plastic material impacted from above by an arbitrary-shaped die. The governing equation for solid material dynamics is given by the equilibrium equation:
\begin{equation}\label{eq:solid}
    \rho_s \frac{\partial^2 u}{\partial t^2} + \nabla \cdot \sigma = 0,
\end{equation}
where $\rho_s$ represents the solid density, $u$ denotes the displacement vector of the material over time $t$, and $\sigma$ is the stress tensor. The input is the die shape recorded on a $101 \times 31$ structured mesh, and the output is the deformation of each mesh point over the future 20 time steps.

\paragraph{Navier-Stokes.} This dataset simulates incompressible, viscous flow on a unit torus. Since the density of the fluid is constant, the energy conservation is independent, and the fluid dynamics are simplified to the vorticity formulation:
\begin{align}\label{eq:ns}
    \nabla \cdot U &= 0, \\
    \frac{\partial \omega}{\partial t} + U \cdot \nabla \omega &= \nu \nabla^2 \omega + f,
\end{align}
where $U = (u, v)$ is the velocity vector, $\omega = \nabla \times U$ is the vorticity, and the viscosity is set to $\nu = 10^{-5}$. We predict the future 10 frames based on the past 10 frames on a $64 \times 64$ regular grid.

\paragraph{Darcy Flow.} This benchmark models fluid diffusion through porous media, governed by the 2D steady-state elliptic equation:
\begin{equation}\label{eq:darcy}
    -\nabla \cdot (a(x) \nabla u(x)) = f(x),
\end{equation}
where $a(x)$ is the permeability coefficient field. This dataset tests the model's ability to resolve high-frequency spatial discontinuities in the coefficient field $a(x)$ on an $85 \times 85$ regular grid.

\paragraph{Elasticity.} These benchmarks estimate the inner stress of an incompressible material with an arbitrary central void under external tension. They share the same governing equilibrium equation Eq. (\ref{eq:solid}) as the Plasticity dataset.  The model maps the material structure to the inner stress.

\hspace{0.02in}

\subsection{Irregular domain benchmarks}
\label{eq2}
\paragraph{Irregular Darcy.} This 2D fluid problem is discretized with 2,290 unstructured nodes. It shares the same 2D steady-state elliptic equation as the standard Darcy flow Eq.~(\ref{eq:darcy}).
However, it challenges the model with highly irregular geometric boundaries. The task maps the varying permeability coefficient field $a(x)$ to the pressure field $u(x)$.

\paragraph{Pipe Turbulence.} Discretized with 2,673 nodes, this scenario requires state predictions of incompressible fluid dynamics within an unstructured pipe geometry. The governing equations are the as before Eq.~(\ref{eq:pipe}).
The model is tasked with predicting the future velocity states based on previous observations in the irregular domain.

\paragraph{Composite.} We scale our evaluation to a challenging 3D solid mechanics task using the Composite material benchmark (8,232 nodes). This dataset models the thermoelastic deformation of a 3D composite material driven by temperature profiles. The underlying physics is governed by the steady-state equilibrium equation coupled with thermal expansion:
\begin{equation}
    \nabla \cdot \sigma = 0, \quad \text{where} \quad \sigma = \mathbf{C} : \varepsilon - \gamma \Delta T \mathbf{I}.
\end{equation}
Here, $\sigma$ is the stress tensor, $\mathbf{C}$ is the stiffness tensor, $\varepsilon$ is the strain tensor, and $\gamma \Delta T \mathbf{I}$ represents the thermal stress induced by the temperature change $\Delta T$. The model learns to map the input 3D temperature profiles directly to the structural stress fields.

% \noindent {\bf References}

\end{document}